\documentclass[10pt,letterpaper]{article}
\usepackage{opex3}
\usepackage{color}
\begin{document}

%%%%%%%%%%%%%%%%%% title page information %%%%%%%%%%%%%%%%%%

\title{Length sensing and control of a Michelson interferometer with Power Recycling and Twin Signal Recycling cavities}

\author{Christian Gr\"{a}f, Andr\'{e} Th\"{u}ring, Henning Vahlbruch,\\ Karsten Danzmann and Roman Schnabel}

\address{Institut f\"{u}r Gravitationsphysik, Leibniz Universit\"{a}t Hannover and Max-Planck-Institut f\"{u}r Gravitationsphysik (Albert-Einstein Institut), Callinstra\ss{}e 38, 30167 Hannover, Germany}

\email{christian.graef@aei.mpg.de} %% email address is required

%% \homepage{http:...} %% author's URL, if desired

%%%%%%%%%%%%%%%%%%% abstract and OCIS codes %%%%%%%%%%%%%%%%
%% [use \begin{abstract*}...\end{abstract*} if exempt from copyright]

%%%%%%%%%%%%%%%%%%%%%%%%%%  body  %%%%%%%%%%%%%%%%%%%%%%%%%%

\begin{abstract}
The techniques of power recycling and signal recycling have proven as key concepts to increase the sensitivity of 
large-scale gravitational wave detectors by independent resonant enhancement of
light power and signal sidebands within the interferometer.
Developing the latter concept further, \emph{twin signal recycling} 
was proposed as an alternative
to conventional detuned signal recycling.
Twin signal recycling features the narrow-band 
sensitivity gain of conventional detuned signal recycling but furthermore facilitates the injection of squeezed states of light, increases 
the detector sensitivity over a wide frequency band 
and requires a less complex detection scheme for 
optimal signal readout.
These benefits come at the expense of an additional recycling mirror, 
thus increasing the number of degrees of freedom in the interferometer which need to be controlled.

In this article we describe the development of a length sensing and control
scheme and its successful application to a tabletop-scale 
power recycled Michelson interferometer with twin signal recycling. 
We were able to lock the interferometer in all
relevant longitudinal degrees of freedom, 
enabling the long-term stable operation of the experiment. 
We thus laid the foundation for further investigations of this interferometer topology to evaluate its viability 
for the application in gravitational wave detectors.
\end{abstract}

 % REPLACE WITH CORRECT OCIS CODES FOR YOUR ARTICLE

%%%%%%%%%%%%%%%%%%%%%%% References %%%%%%%%%%%%%%%%%%%%%%%%%

%%%%%%%%%%%%%%%%%%%%%%%%%%%%%%%%%%%%%%%%%%%%%%%%%%%%%%%%%%%%%%%%%%%%%%%%%%%%%%%%%%%%%%%%%%%%%%%%%%%%%%%%%%%%%%%%%%%%%%%%%%%%%%%%%%%%%%%%%%%%%%%%%%%%%%%%%%%%%%%%%
%%%%%%%%%%%%%%%%%%%%%%%%%%%%%%%%%%%%%%%%%%%%%%%%%%%%%%%%%%%%%%%%%%%%%%%%%%%%%%%%%%%%%%%%%%%%%%%%%%%%%%%%%%%%%%%%%%%%%%%%%%%%%%%%%%%%%%%%%%%%%%%%%%%%%%%%%%%%%%%%%
\section{Introduction}
The international network of large-scale interferometric gravitational wave (GW) 
detectors is currently undergoing an extensive technological upgrade towards what 
is commonly referred to as the ``second generation'' of GW detectors. Upon completion, 
these second generation observatories, namely GEO-HF \cite{willke06}, Advanced LIGO 
\cite{AdvancedLIGO} and Advanced Virgo \cite{AdvancedVirgo}, complemented by the new 
observatory KAGRA \cite{KAGRA}, will reach unprecedented sensitivities to
GW-induced 
strain. This sensitivity improvement constitutes a major leap towards reaching the 
long-standing goal of the first direct measurement of GWs.

Besides power recycling (PR), which was already implemented in the first generation 
LIGO and Virgo interferometers, signal recycling \cite{meers88} (SR) has been adopted 
for the optical layouts of all second generation detectors. Dual recycling, i.e.~the 
combination of PR and SR, has already been successfully implemented and operated in 
the GEO\,600 detector \cite{hild07}.

Furthermore, the injection of squeezed vacuum -- a technique proposed by Caves 
to improve the quantum noise limited sensitivity of GW detectors \cite{caves81} 
-- will be employed to enhance the performance of GEO-HF as well as the Advanced 
LIGO Hanford observatory. The current GEO\,600 detector has been the first 
large-scale interferometer to successfully demonstrate long-term stable squeezing 
injection and the associated broadband sensitivity improvement, already for the 
case of carrier-tuned SR. \cite{nature_sqz}

The compatibility of \emph{detuned} SR interferometers with squeezing injection 
was investigated theoretically by Harms et al.~\cite{harms03}. This work led to 
the insight that the combination of these two techniques requires additional 
filter cavities in the squeezing path to take full advantage of the benefit from 
squeezing injection on the detector sensitivity. The use of filter cavities was 
previously suggested by Kimble et al.~to compensate for radiation pressure noise 
in GW detectors \cite{kimble02}. 

\begin{figure}[t]
\centering
\includegraphics[width=0.8\textwidth, clip=true, trim=0cm 0cm 0cm 0cm]{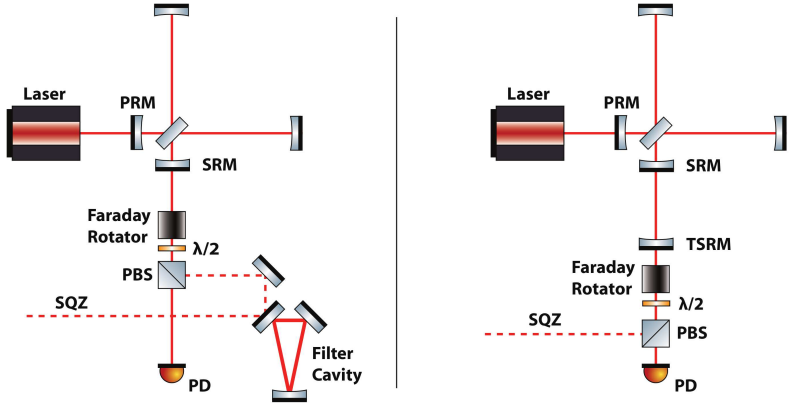}
\caption{General comparison of the dual recycling (left) and the power recycled 
twin signal recycling Michelson (right) topologies, both including squeezed light 
injection. Compared to dual recycling, the power recycled TSR interferometer
features a second recycling mirror in the detection port, TSRM, which forms a 
three mirror coupled cavity together with the SRM and the Michelson end mirrors. 
For a broadband sensitivity improvement at shot noise limited frequencies by 
squeezing injection, the dual recycling interferometer with detuned SR requires 
an additional filter cavity, to compensate for the SR cavity-induced rotation of 
the squeezing ellipse. Contrasting this, the TSR topology is inherently compatible 
with squeezing at shot noise limited frequencies and requires no additional optical 
filter in the squeezing path.}
\label{fig:drmi_tsr_topologies}
\end{figure}

As an alternative to conventional detuned SR, Th\"{u}ring et al.~proposed the technique 
of twin signal recycling \cite{thuering07} (TSR) which is based on introducing a second 
recycling mirror in the asymmetric port of the interferometer (cf.~Fig.~\ref{fig:drmi_tsr_topologies}). 
It was shown that, in contrast to detuned SR, an interferometer featuring TSR does not 
require additional filter cavities to maximize the benefit from squeezing injection in 
the shot noise-limited frequency regime. Besides this, Th\"{u}ring points out
further 
advantages of TSR over conventional detuned SR in his paper: a sensitivity improvement 
over a wide frequency band and a less challenging GW signal readout scheme. The former 
can be attributed to the simultaneous resonant enhancement of both the upper and the 
lower GW signal sidebands, which can be arranged for in TSR configurations. 
The latter stems from the fact that in TSR configurations the GW signal is entirely 
contained in the phase quadrature of the carrier field, similar to the case of a
conventional Michelson interferometer.

In this paper we describe the electronic stabilization and long-term operation
of a 
power recycled Michelson interferometer with TSR, thus laying the foundation for further 
investigations of this topology to evaluate its viability for the application in GW 
interferometry. One such aspect, the broadband enhancement of the interferometer by 
squeezed light injection, was carried out in a subsequent experiment and was already 
reported on in \cite{thuering09}.

%%%%%%%%%%%%%%%%%%%%%%%%%%%%%%%%%%%%%%%%%%%%%%%%%%%%%%%%%%%%%%%%%%%%%%%%%%%%%%%%%%%%%%%%%%%%%%%%%%%%%%%%%%%%%%%%%%%%%%%%%%%%%%%%%%%%%%%%%%%%%%%%%%%%%%%%%%%%%%%%%
%%%%%%%%%%%%%%%%%%%%%%%%%%%%%%%%%%%%%%%%%%%%%%%%%%%%%%%%%%%%%%%%%%%%%%%%%%%%%%%%%%%%%%%%%%%%%%%%%%%%%%%%%%%%%%%%%%%%%%%%%%%%%%%%%%%%%%%%%%%%%%%%%%%%%%%%%%%%%%%%%
\section{Layout of the power recycled TSR interferometer} \label{sec:optical_layout_of_the_tsr_experiment}
The starting point of our investigations was an existing dual recycling Michelson interferometer (DRMI) 
tabletop experiment \cite{vahlbruch05} -- a scaled and largely simplified model of 
the GEO\,600 interferometer -- which we were aiming to extend to a power
recycled TSR interferometer with a minimum of invasive changes. Thus, we were able 
not only to experimentally test the power recycled TSR topology but also to evaluate 
the viability of TSR as an upgrade for existing interferometers with conventional SR.

\subsection{TSR cavity parameters} \label{subsec:optical_design_parameters}
By introducing an additional recycling mirror into the optical setup of the DRMI, the former 
signal recycling cavity (SRC) 
is converted into a system of two optically coupled cavities which we refer to
as the \emph{twin signal recycling cavity} (TSRC) throughout this paper.
This cavity is formed by the newly introduced mirror and the Michelson mirrors 
acting as the end mirrors, and the former signal recycling mirror acting as the coupling mirror. 
For clarity we will keep referring to the former signal recycling mirror as SRM 
and denote the newly introduced mirror as \emph{twin signal recycling mirror} (TSRM). 
We will refer to the length of the TSRM--SRM cavity
as $L_{\textnormal{\small{SR1}}}$ and, correspondingly, the average length of the cavity 
formed by the SRM and the Michelson end mirrors (i.e.~the former SRC) will be denoted as
$L_{\textnormal{\small{SR2}}}$, in accordance with the nomenclature used in \cite{thuering07}.

Three free parameters of the TSRM were to be determined: (i) reflectivity, 
(ii) radius of curvature (ROC) and (iii) 
distance to the SRM. These, either directly or indirectly, affect the 
frequency response as well as the structure of the sensing matrix of the 
``optical plant'' which determines the complexity of the longitudinal control scheme. 

The reflectivity of the TSRM influences the level of resonant enhancement of the signal 
sidebands, which is in close analogy to the role of the SRM reflectivity in a signal 
recycled interferometer.
We chose a reflectivity of $R_{\textnormal{\small{TSRM}}}=95$\% for our experiment.

The ROC of the TSRM and its distance to the SRM can, generally, not be chosen 
independently. The boundary conditions for a stable and (theoretically) well 
mode matched cavity constrain the values these parameters can reasonably take on. 
From the control point of view it was 
desirable to chose $L_{\textnormal{SR1}}$ such that, for given phase modulation (PM) control 
sideband frequencies, an  error signal could be extracted for feedback 
to the TSRM which preferably exhibits only small coupling to the other 
longitudinal degrees of freedom (DOF) in the interferometer. For a known
length $L_{\textnormal{SR1}}$ the corresponding TSRM curvature, which results 
in a stable and mode matched cavity, can then be easily determined.
We will discuss our approach of determining the optimal length 
$L_{\textnormal{\small{SR1}}}$, driven by control considerations, in the following section.

\subsection{Development of a sensing scheme} \label{subsec:development_of_a_sensing_scheme}
For the design of the sensing scheme we performed a series of numerical 
simulations, based on a steady-state frequency domain model of the power
recycled TSR interferometer realized with the software \emph{Finesse} 
\cite{freise04}. 
Besides the length $L_{\textnormal{SR1}}$ the optimal 
locations of the length signal extration ports to obtain error signals for the 
four longitudinal DOF of the interferometer were
to be determined with the aid of the numerical model. These DOF are the differential mode of the 
Michelson arms (MICH), the power recycling cavity (PRC) and the microscopic 
positions of the TSRM relative to the SRM ($\delta \textnormal{L}_{\small{\textnormal{SR1}}}$) 
and the SRM relative to the average position of the Michelson end mirrors
($\delta \textnormal{L}_{\small{\textnormal{SR2}}}$), respectively.

\begin{figure}[h]
\centering
   \includegraphics[width=0.85\textwidth]{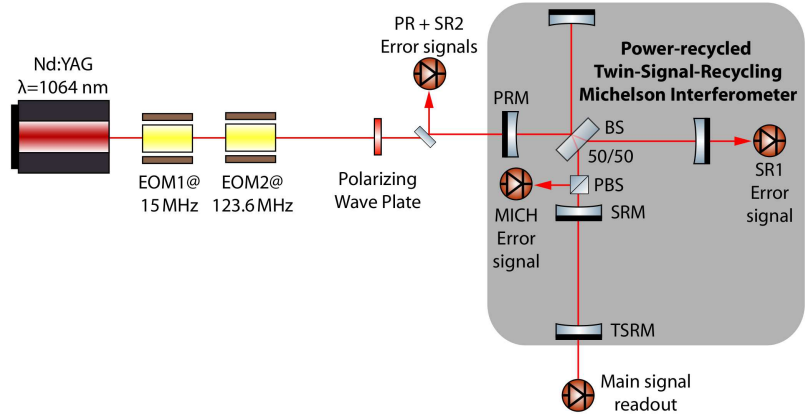}
\caption{Schematic drawing of the conceptual approach to length sensing in the TSR interferometer. The sensing scheme is based on two different pairs of phase modulation sidebands and four 
heterodyne signal extraction ports to obtain four error signals for feedback control of the longitudinal DOF of the interferometer. The wave plate in the input beam path and the polarizing
beam splitter between the Michelson beam splitter and the SRM permit interferometer locking with orthogonal polarization modes.}
\label{fig:signal_extraction_ports}
\end{figure}
 
Due to the coupled nature of the longitudinal DOF in a power recycled TSR interferometer 
the extracted heterodyne length signals, likewise, are coupled. 
This coupling can often be reduced by prudently designing the optical plant and the sensing system.
However, the RF modulation scheme was carried over from the precursor DRMI experiment and modulation frequencies were thus
not intended as free parameters for the optimization of the sensing scheme. 
As a consequence, the resulting sensing matrices were non-orthogonal but could
 be arranged for to have at least full rank by choosing appropriate values for the
remaining free parameters. In this case the technique of \emph{gain hierarchy} 
\cite{strain03} can often be applied and was successfully used in earlier 
interferometry experiments, e.g.~in \cite{regehr95}. 

\begin{table}[ht]
  \caption{Modulation/demodulation parameters of the signal extraction system. 
In the first column the longitudinal DOF are listed along with, in brackets, 
the mirrors which were actuated upon to tune the respective DOF.}
  \centering
  \begin{tabular}{l|c|c|c}
  \hline
  DOF (Mirror) & $f_{\textnormal{\tiny{demod}}}$ & $\phi_{\textnormal{\tiny{demod}}}$ & Detection port \\
  \hline\hline
  PRC (PRM) & 15\,MHz & in-phase & PRC reflected beam  \\
  MICH (EMx)& 123.6\,MHz & quadrature & internal pick-off (p-pol)\\
  $\delta \textnormal{L}_{\small{\textnormal{SR1}}}$ (TSRM) & 123.6\,MHz & in-phase & MI end mirror trans. beam \\
 $\delta \textnormal{L}_{\small{\textnormal{SR2}}}$ (SRM) & 123.6\,MHz & in-phase & PRC reflected beam \\
  \hline
  \end{tabular}
  \label{tab:signal_extraction_params}
\end{table}

To facilitate lock acquisition of the interferometer, polarization
dependent sensing was employed to decouple length signals of coupled DOF.
In this approach the polarization of the input beam is chosen to be
a linear combination of s- and p-polarized light. A polarizing beam splitter 
between the Michelson beam splitter and the SRM serves to separate s-polarized 
light, which circulates in the TSRC, from p-polarized light, which exclusively 
senses the MICH degree of freedom. Thus, the error signal extracted from the p-polarized field 
can be used to lock the MICH degree of freedom and, furthermore, it exhibits little 
sensitivity to perturbations in the other longitudinal DOF.

\noindent The simulations resulted in the following findings:
\begin{itemize}

\item The signal extraction ports in the optical setup were chosen with respect to optimal decoupling of all heterodyne length signals.
Our preferred signal extraction port configuration is depicted in Fig. \ref{fig:signal_extraction_ports}, further details are given in Tab. \ref{tab:signal_extraction_params}.  

\item For the general case of unequal lengths $L_{\textnormal{\small{SR1}}} \neq L_{\textnormal{\small{SR2}}}$ of the TSRC, the longitudinal degrees of freedom MICH and PRC showed strong coupling to the error signals for the SRM and TSRM positions.

 \item We found that in order to reduce the coupling of MICH and the SRM position to the TSRM error signal
it was desirable to arrange for identical lengths in the TSRC, i.e.~choosing $L_{\textnormal{\small{SR1}}}=L_{\textnormal{\small{SR2}}}$.

\item Finally, from the optimized sensing matrix given in Tab. \ref{tab:sensing_matrix_num} we could deduce a suitable order for the hierarchical locking of the interferometer. Lock acquisition starts with (i) the PRC, followed by (ii) MICH, (iii), the TSRM, and is completed with the stabilization of (iv) the SRM position.
\end{itemize}

\begin{table}[ht]
  \caption{Normalized sensing matrix as obtained from the numerical model of the TSR interferometer. Rows correspond to different signal extraction ports, columns to different longitudinal DOF. For the calculation
of each entry it was assumed implicitly that the other DOF were at their designated operating points.
Heterodyne length signals which stem from unwanted coupling of neighboring DOF are highlighted in blue, 
numerical zeros are printed in green.}
  \centering
  \begin{tabular}{l|c|c|c|c}
  \hline
   & PRC & MICH & $\delta \textnormal{L}_{\small{\textnormal{SR2}}}$ & $\delta \textnormal{L}_{\small{\textnormal{SR1}}}$ \\
  \hline\hline
  PD$_\textnormal{PRC}$ & $1$ & $1.18 \times 10^{-6}$ & \color{green}$0$\color{black} & \color{green}$0$\color{black} \\
  PD$_\textnormal{MICH}$ & $-9.32 \times 10^{-4}$ & 1 & 0 & 0 \\
  PD$_\textnormal{SR2}$ & \color{blue}{$-579.19$}\color{black} & \color{blue}{$-32.18$}\color{black} & 1 & \color{blue}{$-55.70$}\color{black} \\
  PD$_\textnormal{SR1}$ & \color{blue}{$-3.07$}\color{black} & $-0.12$ & $8.18 \times 10^{-3}$ & $1$ \\
  \hline
  \end{tabular}
  \label{tab:sensing_matrix_num}
\end{table}

%%%%%%%%%%%%%%%%%%%%%%%%%%%%%%%%%%%%%%%%%%%%%%%%%%%%%%%%%%%%%%%%%%%%%%%%%%%%%%%%%%%%%%%%%%%%%%%%%%%%%%%%%%%%%%%%%%%%%%%%%%%%%%%%%%%%%%%%%%%%%%%%%%%%%%%%%%%%%%%%%
%%%%%%%%%%%%%%%%%%%%%%%%%%%%%%%%%%%%%%%%%%%%%%%%%%%%%%%%%%%%%%%%%%%%%%%%%%%%%%%%%%%%%%%%%%%%%%%%%%%%%%%%%%%%%%%%%%%%%%%%%%%%%%%%%%%%%%%%%%%%%%%%%%%%%%%%%%%%%%%%%
\section{Experimental setup} \label{sec:experimental_setup}
\begin{figure}[t]
\centering
   \includegraphics[width=\textwidth]{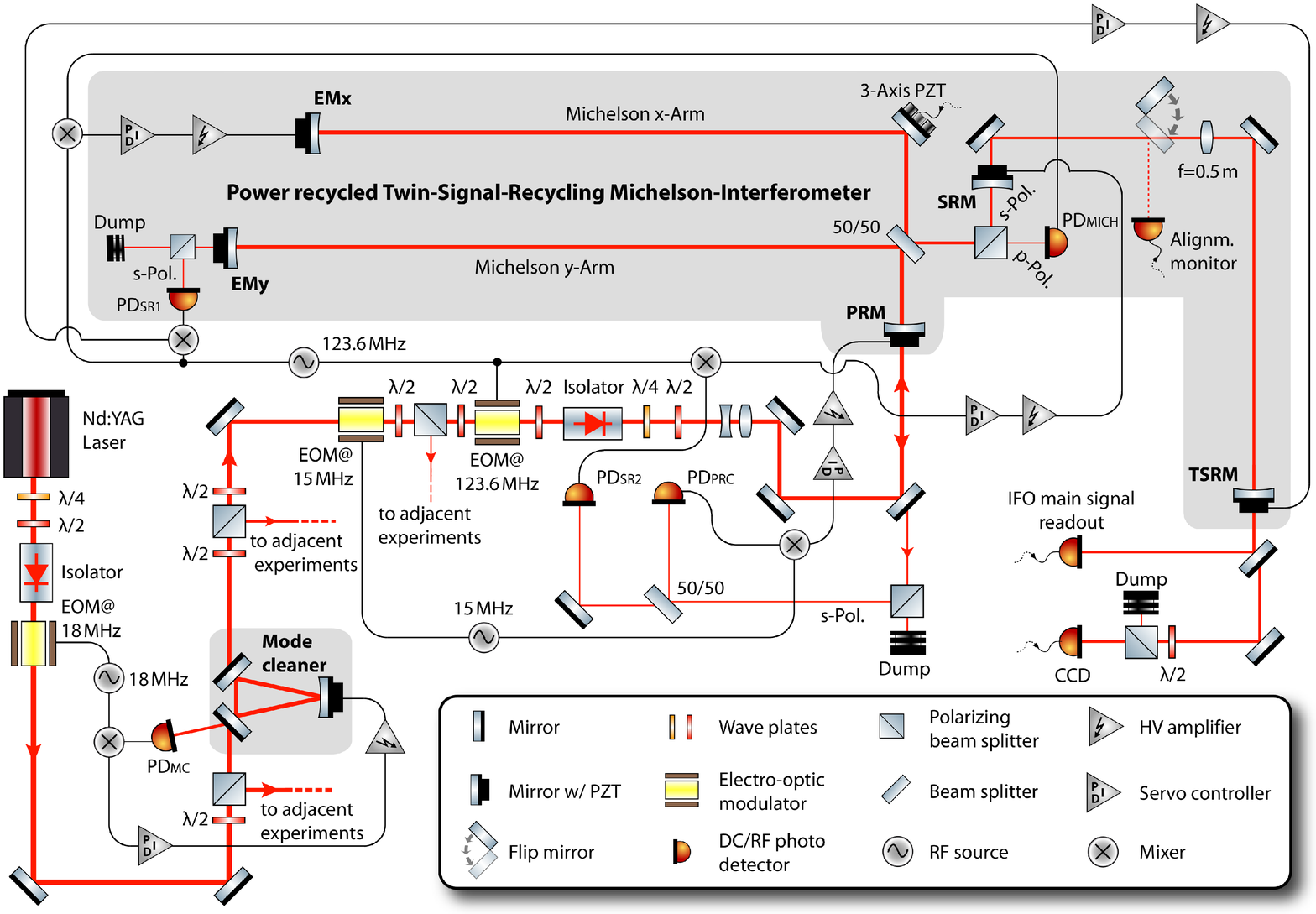}
\caption{Optical layout of the power recycled twin signal recycling interferometer experiment. A number of auxiliary optics were omitted for clarity.}
\label{fig:optical_layout}
\end{figure}

We now proceed with a discussion of the optical setup of the interferometer, depicted in the schematic drawing in Fig.~\ref{fig:optical_layout}.

The main laser source was a Nd:YAG Innolight Mephisto NPRO laser with a nominal cw output power of 2\,W at 1064\,nm which was shared with adjacent experiments. 
Approximately 50\% of the 
light was picked off and transmitted through a rigid, aluminum spacer-mounted ring cavity (``mode cleaner'') for spatial filtering and the suppression of high-frequency amplitude and phase noise of the laser beam.
The PDH 
reflection locking technique \cite{drever83} was employed to keep the mode cleaner cavity (MC) on resonance. 
Control sidebands were imprinted on the laser field by directing the beam through an electro-optic modulator (EOM) which was driven
by an electronic oscillator at 18\,MHz.  
Appropriate feedback signals were extracted by demodulating the field reflected 
by the MC, to obtain the 18\,MHz beat signal of the PM sidebands with the carrier. 
A piezoelectric transducer (PZT), which was glued to one 
of the MC cavity mirrors, served as an actuator to tune the cavity length. The MC was locked by applying 
a high voltage (HV) amplified correction signal to the PZT, which was generated by servo electronics.

Prior to its injection into the interferometer the filtered laser beam was directed through 
two EOMs to imprint two pairs of PM control sidebands on the input field, at 15\,MHz and at 123.6\,MHz, respectively. 
The EOMs were followed by a Faraday isolator to attenuate back-reflected light from the interferometer and, in conjunction 
with a wave plate, to adjust the laser power at the interferometer input. 
Another two wave plates were placed between the isolator and the PRM: 
a quarter wave plate to eliminate elliptically polarized components of the input field and a half wave plate to adjust the linear polarization vector of the input beam.
Together with the polarizing beam splitter (PBS) in the interferometer, between the main beam splitter (BS) and the SRM, 
this allows for independent sensing of the two polarization modes, thus providing an effective way to obtain a signal for MICH which exhibited
nearly perfect decoupling from the position of the SRM (cf.~Sec.~\ref{subsec:development_of_a_sensing_scheme}).
A number of lenses and steering mirrors in the input optics optical train served to match the input beam to the eigenmode of the PRC. 
For the investigations discussed in this paper the interferometer was illuminated with approximately 40\,mW of input laser light, composed of 75\% s-polarized and 25\% p-polarized light.

The beam transmitted by the PRM, which had a power reflectivity of $R_{\textnormal{\small{PRM}}}=90$\% and a ROC of $2$\,m, propagated towards the BS of the Michelson interferometer where it
was split into two beams of equal power which then propagated along the parallel (due to space limitations) interferometer arms.
The parallel arrangement of the arms required a steering mirror in the x-arm which was mounted in 3-axis piezo-driven mirror mount for easy ``on-the-fly'' tuning of the interferometer alignment. 
Both interferometer end mirrors (EMx, EMy) had a nominal ROC of 1.5\,m and a power reflectivity of 99.92\%. A Schnupp
asymmetry of 7\,mm was introduced to the Michelson arms to allow for 
control sideband leakage into the detection port while the interferometer was locked on the dark fringe \cite{mizuno95}.

The two recycling mirrors in the asymmetric port were placed at distances of 0.21\,m (SRM) and 1.42\,m (TSRM) from the Michelson BS, respectively, measured with respect to their reflective surfaces. 
Reflectivity and curvature of the SRM were 90\% and 2\,m, respectively. The plane TSRM was coated for a reflectivity of 95\%.
For the cavity formed by the SRM and the TSRM a convex-plane geometry was
adopted, with an additional plano-convex intra-cavity lens with a focal length
of 0.5\,m. Due to its higher flexibility this configuration was preferred to a TSRM with custom-made ROC.

All five interferometer mirrors were mounted in 3-axis mounts, with a clamped PZT for actuation in the direction perpendicular to the mirror surface. 
The positions of the PRM, the SRM and the TSRM were actuated on by applying a HV signal to their respective PZT actuator. Actuation of the MICH degree of freedom was achieved 
by driving a PZT in one of the Michelson end mirror mounts.
  
Numerous photo detectors were employed to extract RF signals for length sensing as well as DC signals for diagnostics, e.g.~to 
determine the quality of the alignment and mode matching of the experiment and
for monitoring of the ``state'' of each of the longitudinal DOF (either locked or freely fluctuating). 
A CCD camera was 
placed in transmission of a steering mirror in the detection port for online monitoring of the spatial shape of the beam leaving the interferometer. Furthermore a number of flip-mirrors were introduced in the setup, to pick off light internally for alignment and diagnostics purposes.

Due to the optical coupling of the Michelson interferometer with the recycling cavities it was a non-trivial task to bring the interferometer to a well-aligned state. To optimize 
the angular alignment and the mode matching of the interferometer, an iterative alignment scheme was developed. 
A necessary prerequisite was to ``disable'' the PRC prior to optimizing the Michelson fringe contrast and aligning the TSRC.
This suppression of the PRC resonance was achieved by tilting the PRM. The slightly different optical path lengths of the tilted and the aligned PRM had no observable degrading effect on the final 
alignment quality. 

\begin{table}[ht]
  \caption{Design parameters of the power recycled TSR interferometer experiment.}
  \centering
  \begin{tabular}{ll|ll}
  \hline\hline
  Michelson arm length & 1\,m $\pm$ 7\,mm & Nominal ROC EMx, EMy & 1.5\,m \\
  Distance of PRM/SRM to BS & 0.21\,m & Nominal ROC PRM/SRM & 2\,m\\
  Distance SRM to TSRM ($L_{\textnormal{\small{SR1}}}$) & 1.21\,m & Nominal ROC TSRM & plane \\
  Recycling cavity FSR & 123.6\,MHz & Intra-cavity lens focal length& 0.5\,m\\
Reflectivity EMx, EMy & 99.92\% & Beam waist size in PRC & 499\,\textmu m\\
Reflectivity PRM/SRM, TSRM & 90\%, 95\% & Beam waist size on TSRM & 181\,\textmu m\\
  \hline\hline
  \end{tabular}
  \label{tab:exeriment_params}
\end{table}

%%%%%%%%%%%%%%%%%%%%%%%%%%%%%%%%%%%%%%%%%%%%%%%%%%%%%%%%%%%%%%%%%%%%%%%%%%%%%%%%%%%%%%%%%%%%%%%%%%%%%%%%%%%%%%%%%%%%%%%%%%%%%%%%%%%%%%%%%%%%%%%%%%%%%%%%%%%%%%%%%
%%%%%%%%%%%%%%%%%%%%%%%%%%%%%%%%%%%%%%%%%%%%%%%%%%%%%%%%%%%%%%%%%%%%%%%%%%%%%%%%%%%%%%%%%%%%%%%%%%%%%%%%%%%%%%%%%%%%%%%%%%%%%%%%%%%%%%%%%%%%%%%%%%%%%%%%%%%%%%%%%
\section{Lock acquisition and measurements}
In this section we 
discuss the aspect of bringing the aligned and mode matched, yet uncontrolled interferometer to a fully locked state of its four longitudinal DOF.

\begin{figure}[th]
\centering
   \includegraphics[width=0.97\textwidth]{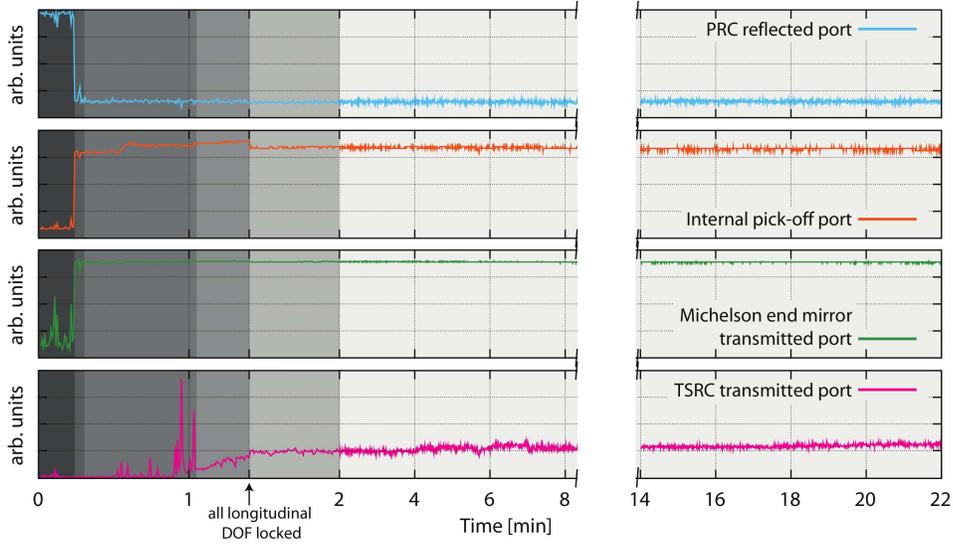}
\caption{Monitored photo detector DC transients, recorded during lock acquisition and fully locked operation of
the interferometer. The traces show light powers measured in the symmetric port (i.e. in reflection of the
power recycling cavity, blue trace), in the Michelson-internal pick-off port (orange trace), in the port in transmission
one of the Michelson end mirrors (green trace) and in the asymmetric port (i.e. in transmission of the TSRC, magenta
trace). 
The grey shaded areas represent different stages of the acquisition sequence, resulting in a fully locked interferometer after $t=1.4$\,min. The proportional gains of the feedback loops were optimized for long-term stable operation at $t=2$\,min.}
\label{fig:dc_levels_locking}
\end{figure}

\begin{figure}[th]
\centering
   \includegraphics[width=0.97\textwidth]{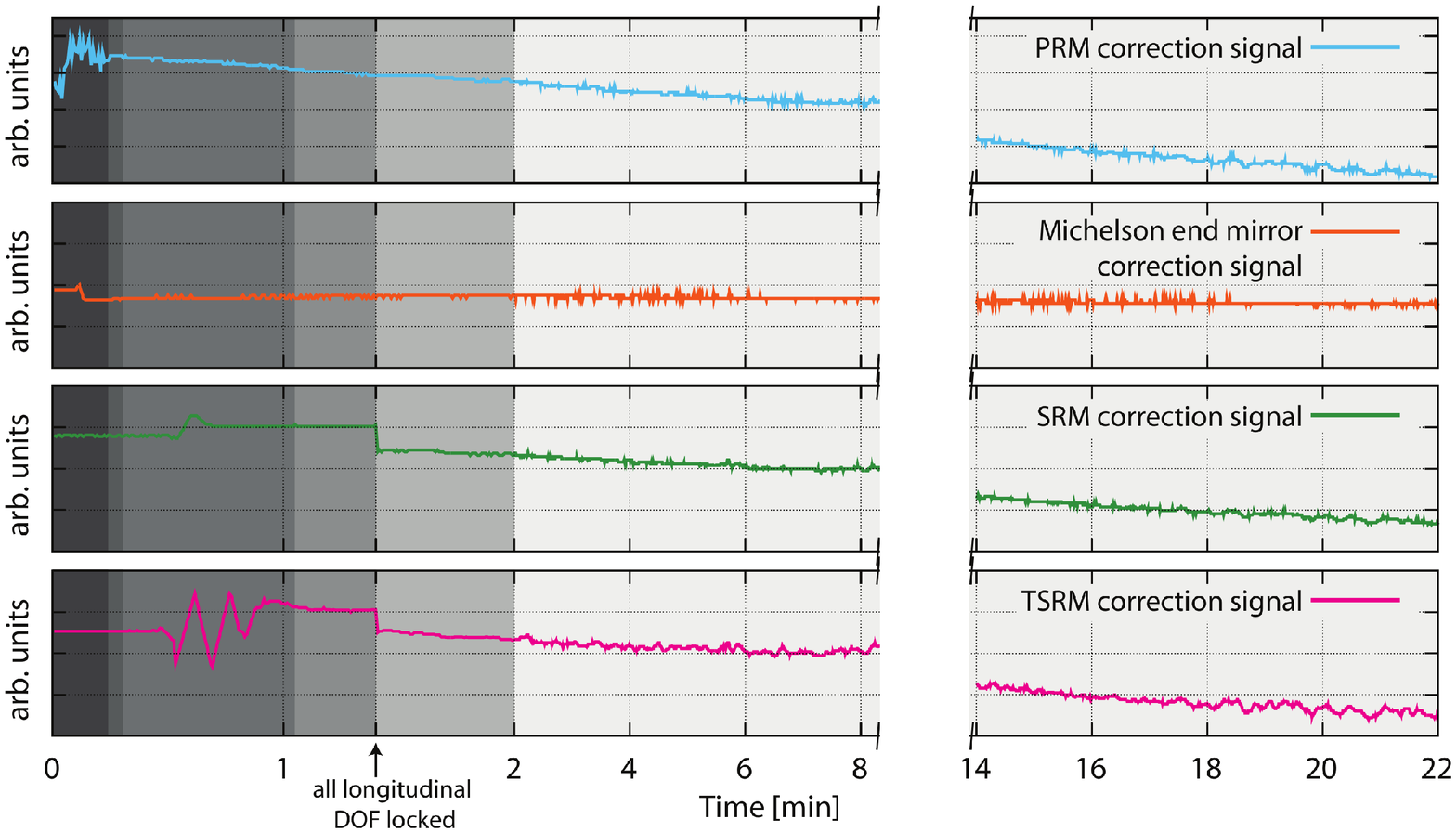}
\caption{Monitored correction signals generated by the servo controllers for the four longitudinal DOF. The traces were recorded simultaneously with the data shown in Fig. \ref{fig:dc_levels_locking}. 
These signals were amplified and in turn applied to piezoelectric transducers actuating on the power recycling mirror
(blue trace), one of the Michelson end mirrors (orange trace), the signal recycling (green trace) and the 
twin signal recycling mirror (magenta trace), respectively. Long term drifts eventually led to a lock-loss of the
instrument due to the limited range of the actuators.}
\label{fig:control_signals_locking}
\end{figure}

The hierarchic stabilization of the lengths in the 
interferometer was carried out in the order which was derived from the numerical interferometer model, as described in Sec. \ref{subsec:development_of_a_sensing_scheme}.
Due to different types of imperfections, which inevitably occur in any real interferometer, the extracted heterodyne signals were likely to deviate from those obtained from the strongly idealized
numerical model. Among these imperfections are e.g.~non-ideal mode matching and the presence of higher-order modes, non-optimal demodulation phases, deviations from nominal lengths and
also the coupling of length fluctuations from yet unstabilized DOF, which was not included in the model. Several iterations were necessary to optimize the relevant parameters and 
settings to fully lock the interferometer for the first time. In this stage
of the experiment the CCD camera in the detection port helped to determine whether the interferometer was accidentally locked  on some residual higher order mode or on the 
desired fundamental mode.

As a figure of merit, and also to document the transition of the interferometer from the initial, unlocked state to a fully locked final state, light powers in 
different optical ports along with the correction signals generated by the servo controllers were monitored during the acquisition phase and after lock was acquired. 
We used two digital storage oscilloscopes to record the data of two consecutive exemplary measurements which we present
in Figs.~\ref{fig:dc_levels_locking} and \ref{fig:control_signals_locking}. 

The shaded regions in the plots represent different stages of the lock acquisition process:

\begin{enumerate}
 \item All DOF unlocked, mirror positions randomly fluctuating.
 \item MICH manually tuned close to the dark fringe, PRC length locked (red traces).
 \item MICH locked on the dark fringe using p-polarized light (green traces).
 \item The TSRM position was scanned by applying a triangular drive to its PZT. The SRM was tuned manually to its operating point to obtain a decent error signal for feedback to the TSRM, to lock $\delta \textnormal{L}_{\textnormal{SR1}}$ (yellow traces).
 \item The position of SRM stabilized to its designated operating point, $\delta \textnormal{L}_{\textnormal{SR2}}$ locked (blue traces).
 \item All longitudinal DOF are electronically stabilized. Proportional gains of the servo loops adjusted to optimize loop suppression.
\end{enumerate}
In Fig.~\ref{fig:control_signals_locking}, when $\delta \textnormal{L}_{\textnormal{SR2}}$ lock is acquired at 1.4 minutes, we can observe a correlated jump in the SRM and TSRM correction signals. This is due to the attempt of the
TSRM servo to maintain the operating point for $\delta \textnormal{L}_{\textnormal{SR1}}$, i.e.~the servo commands
the TSRM to swiftly follow the SRM when the SRM position is locked.

The p-polarization-based probing of the MICH degree of freedom gave rise to a
small amount of s-polarized light leaking out of the asymmetric port. 
This is due to a non-degeneracy of s- and p- polarized light with respect to covered
optical path lengths in the interferometer and resulted in a slight offset of the Michelson operating point from the dark fringe for the s-polarized field.

Special care needed to be taken to prevent oscillations to occur in the servo controllers during 
acquisition. Due to mutual dependencies of the control loops, a consequence 
of the coupling of the longitudinal DOF, the proportional gains of the servo controllers needed to 
be set carefully, to thoroughly balance the suppression of disturbances that 
would cause immediate lock-loss versus loop instability. We also found that
slightly tuning the higher
sideband frequency improved the robustness of our locking scheme. This could be attributed to small deviations of the recycling cavity lengths from their nominal values.

Our optimizations of the experiment led to a gradual increase of the time the 
interferometer remained in a fully locked state of up to tens of minutes. The 
limited duration of the locks could be explained by long-term drifts in the 
longitudinal DOF of the interferometer, e.g.~due to thermal expansion of 
components in the setup but also due to drifts of the input laser frequency. 
Drifts exceeding the range of the PZT actuators   
eventually led to lock-loss of the interferometer. In the example 
shown in Fig.~\ref{fig:dc_levels_locking} and Fig.~\ref{fig:control_signals_locking}, 
lock of all four longitudinal DOF was acquired 1.4 minutes after the measurement 
started and lasted for more than 20 minutes until the actuator range of the 
PRM was exceeded. 

A significant improvement of the duty cycle could be expected from operating 
the interferometer in vacuum. This was, however, not within the scope of the 
experiment described here.

\section{Conclusion and outlook}
In this paper we described the design, the implementation and the successful 
operation of a laboratory-scale power recycled Michelson interferometer with 
twin signal recycling.

With the aid of numerical simulations we derived a length sensing and control 
scheme which we successfully applied to the experiment, to stably lock 
the interferometer in its four longitudinal degrees of freedom. After a series 
of iterative optimizations we were able to operate the fully locked interferometer 
on time scales of tens of minutes, limited by long-term drifts in the optical setup 
exceeding the range of the actuators.

Based on the setup discussed in this paper, the aspect of squeezing injection 
into a twin signal recycling interferometer was investigated in a subsequent 
experiment. Results of this work were already published in \cite{thuering09}.

Future work will include the transfer of the twin signal recycling technique to 
other detector topologies, e.g.~Michelson interferometers with Fabry-Perot arm 
cavities, which will form the basis of all second generation as well as the planned 
third generation observatories, e.g.~the Einstein Telescope \cite{ET}.

\section*{Acknowledgements}
This work has been supported by the International Max Planck Research School (IMPRS)
and the cluster of excellence QUEST (Centre for Quantum Engineering and Space-Time
Research).


\begin{thebibliography}{99}

\bibitem{willke06} B.~Willke for the GEO collaboration, ``The GEO-HF project,'' Class.~Quantum Grav.~{\bf 23,} S207--S214 (2006).

\bibitem{AdvancedLIGO} G.~M.~Harry for the LIGO scientific collaborations, ``Advanced LIGO: the next generation of gravitational wave detectors,'' Class.~Quantum Grav.~{\bf 27,} 084006 (2010).

\bibitem{AdvancedVirgo} The Virgo collaboration, ``Status of the Virgo project,'' Class.~Quantum Grav.~{\bf 28,} 114002 (2011).

\bibitem{KAGRA} K.~Somiya for the KAGRA collaboration, ``Detector configuration of KAGRA -- the Japanese cryogenic gravitational-wave detector,'' Class.~Quantum Grav.~{\bf 29,} 124007 (2012).

\bibitem{meers88} B.~J.~Meers, ``Recycling in laser-interferometric gravitational-wave detectors,'' Phys. Rev. D {\bf 38,} 2317--2326 (1988).

\bibitem{hild07} S.~Hild, H.~Grote, M.~Hewitson, H.~L\"{u}ck, J.~R.~Smith, K.~A.~Strain, B.~Willke, and K.~Danzmann, ``Demonstration and comparison of tuned and detuned signal recycling in a large-scale gravitational wave detector,'' Class. Quantum Grav. {\bf 24,} 1513--1523 (2007).

\bibitem{caves81} C.~M.~Caves, ``Quantum-mechanical noise in an interferometer,'' Phys.~Rev.~D {\bf 23,} 1693--1708 (1981).

\bibitem{nature_sqz} R.~Schnabel for the LIGO scientific collaboration, ``A gravitational wave observatory operating beyond the quantum shot-noise limit,'' Nature Physics {\bf 7,} 962--965 (2011).

\bibitem{harms03} J.~Harms, Y.~Chen, S.~Chelkowski, A.~Franzen, H.~Vahlbruch, K.~Danzmann, and R.~Schnabel, ``Squeezed-input, optical-spring, signal-recycled gravitational-wave detectors,'' Phys. Rev. D {\bf 68,} 042001 (2003)

\bibitem{kimble02} H.~J.~Kimble, Y.~Levin, A.~B.~Matsko, K.~S.~Thorne, and S.~P.~Vyatchanin, ``Conversion of conventional gravitational-wave interferometers into quantum nondemolition interferometers by modifying their input and/or output optics ,'' Phys. Rev. D {\bf 65,} 022002 (2002).

\bibitem{thuering07} A.~Th\"{u}ring, R.~Schnabel, H.~L\"{u}ck, and K.~Danzmann, ``Detuned twin signal recycling for ultra-high precision interferometers,'' \ol {\bf 32,} 985 (2007).

\bibitem{thuering09} A.~Th\"{u}ring, C.~Gr\"{a}f, H.~Vahlbruch, M.~Mehmet, K.~Danzmann, and R.~Schnabel ``Broadband squeezing of quantum noise in a Michelson interferometer with Twin-Signal-Recycling,'' \ol {\bf 34,} 824 (2009).

\bibitem{vahlbruch05} H.~Vahlbruch, S.~Chelkowski, B.~Hage, A.~Franzen, K.~Danzmann, and R.~Schnabel, ``Demonstration of a squeezed-light-enhanced Power- and Signal-Recycled Michelson interferometer,'' \prl {\bf 95,} 211102 (2005).

\bibitem{freise04} A.~Freise, G.~Heinzel, H.~L\"uck, R.~Schilling, B.~Willke, and K.~Danzmann, ``Frequency-domain interferometer simulation with higher-order spatial modes,'' Class. Quantum Grav. {\bf 21,} 1067 (2004).

\bibitem{strain03} K.~A.~Strain, G.~M\"uller, T.~Delker, D.~H.~Reitze, D.~B.~Tanner, J.~E.~Mason, P.~A.~Willems, D.~A.~Shaddock, M.~B.~Gray, C.~Mow-Lowry, and D.~E.~McClelland ``Sensing and control in dual-recycling laser interferometer gravitational-wave detectors,'' \ao {\bf 42,} 1244 (2003).

\bibitem{regehr95} M.~W.~Regehr, F.~J.~Raab, and S.~E.~Whitcomb, ``Demonstration of a power-recycled Michelson interferometer with Fabry-Perot arms by frontal modulation,'' \ol {\bf 20,} 1507 (1995).

\bibitem{drever83} R.~W.~P.~Drever, J.~L.~Hall, F.~V.~Kowalski, J.~Hough, G.~M.~Ford, A.~J.~Munley, and H.~Ward ``Laser Phase and Frequency Stabilization Using an Optical Resonator,'' Appl.~Phys.~B {\bf 31,} 97--105 (1983).

\bibitem{mizuno95} J.~Mizuno, ``Comparison of optical configurations for laser-interferometric gravitational-wave detectors,'' PhD Thesis, (Universit\"at Hannover, 1995).

\bibitem{ET} M.~Punturo et al.~``The Einstein Telescope: a third-generation gravitational wave observatory ,'' Class.~Quantum Grav.~{\bf 27,} 194002 (2010).

\end{thebibliography}
\end{document}